\documentclass[letterpaper, 10 pt, conference]{ieeeconf}  %

\IEEEoverridecommandlockouts                              %
\usepackage{amsfonts} %
\usepackage{amsmath}%
\usepackage{amssymb}%
\usepackage{mathtools}%
\usepackage{hyperref}%
\newcommand{\norm}[1]{\left\lVert#1\right\rVert} %

\usepackage{algorithm}%
\usepackage{algpseudocode} %
\usepackage{xspace}%
\newcommand{\ie}{i.\,e.\xspace}%
\usepackage{blindtext} %
\usepackage{siunitx}
\usepackage[short,nocomma]{optidef} %
\usepackage{siunitx}
\usepackage{microtype}

\usepackage{pgfplots}
\usepackage{tikz}
\pgfplotsset{compat=newest}
\usetikzlibrary{plotmarks}
\usetikzlibrary{arrows.meta}
\usepgfplotslibrary{patchplots}

\newcommand\submittedtext{%
  \footnotesize \textcopyright \the\year{} IEEE. Personal use of this material is permitted. Permission from IEEE must be obtained for all other uses, including reprinting/republishing this material for advertising or promotional purposes, collecting new collected works for resale or redistribution to servers or lists, or reuse of any copyrighted component of this work in other works.}

\newcommand\submittednotice{%
\begin{tikzpicture}[remember picture,overlay]
\node[anchor=south,yshift=10pt] at (current page.south) {\fbox{\parbox{\dimexpr0.65\textwidth-\fboxsep-\fboxrule\relax}{\submittedtext}}};
\end{tikzpicture}%
}
\title{\LARGE \bf
Real-time Gaussian Process based  Approximate Model Predictive Trajectory Tracking Control for Autonomous Vehicles}

\author{Alexander Rose$^1$, Lukas Theiner$^{1}$, Rolf Findeisen$^1$%
\thanks{$^1$Control and Cyber-Physical Systems Laboratory, TU Darmstadt, Germany, {\small \{alexander.rose, lukas.theiner, rolf.findeisen\}@iat.tu-darmstadt.de}}%
}

\begin{document}

\maketitle
\thispagestyle{empty}
\pagestyle{empty}

\begin{abstract}
Applying model predictive control on embedded systems remains challenging due to the high computational cost of solving optimal control problems. To address this limitation, computationally efficient Gaussian process approximations of the implicit model predictive control law can be employed. However, for trajectory-tracking applications, the large amount of training data required for successful generalization across distinct reference trajectories poses a significant challenge. To improve data efficiency, we propose to transform the model into curvilinear coordinates around the reference trajectory. Secondly, we use a nominal feedforward component, allowing the Gaussian process to learn only the residual control input, making the approximation of a trajectory-tracking controller feasible. To underline the applicability of the approach, we deploy the controller on a Raspberry Pi in a small-scale vehicle and validate it experimentally. Compared to a model predictive control implementation using real-time iterations, the Gaussian process based approximation computes control inputs about five times faster while achieving similar closed-loop tracking performance. %

\end{abstract}

\section{Introduction}
\submittednotice
In many modern autonomous systems, such as self-driving cars \cite{Bethge_MPCGPHUMAN23,Yu2021_autonomousDriving}, drones \cite{Lefeber2024}, and industrial robots \cite{Graichen_manipulator}, precise and timely decision-making is essential to ensure safety, efficiency, and optimal performance \cite{CCPS_EmbeddedReview}.
These systems must account for dynamically changing environments, physical limitations, and operational constraints while computing control actions in real time.
Model predictive control (MPC)  \cite{Gruene2017} is suited for this purpose, as it formulates control as an optimization problem that explicitly accounts for system dynamics and constraints, ensuring safety and performance.
However, solving the resulting optimization problem in real-time is challenging, particularly for systems with complex dynamics, high-dimensional state spaces, or fast sampling rates \cite{Wang_fastMPC, Limon_efficientmpc}.
This computational burden often makes real-time MPC challenging for embedded systems or applications requiring high-frequency control updates.

Various strategies have been developed to reduce the computational demands of MPC. 
 One line of research focuses on improving optimization algorithms to solve the underlying problem more efficiently  \cite{CCPS_EmbeddedReview}.
For instance, \cite{RTIDiehl} proposes an approach where, at each closed-loop step, only a quadratic program must be solved, while in \cite{Colin_EfficientIP} the authors propose a variant of an interior point method specifically designed for the needs of MPC. 
Another approach to  overcome the computational burden of MPC is to reduce a complex model or cast the optimization problem into a lower dimensional space.
For example, in \cite{Bieker2019DeepMPC} recurrent neural networks are used to predict only the control relevant states for a MPC of a complex fluid system. 
 In \cite{EdgarModelReductionMPC} the authors reduce a nonlinear model using balanced truncation.
 As shown, the closed loop performance of the MPC with a reduced model is similar to the performance of the MPC with the nominal model.
 For linear systems, \cite{Schurig_dimreduction} proposes a method to design low dimensional subspaces for the decision variables, while guaranteeing initial feasibility of the optimization problem.%

A further means of accelerating MPC is to modify the cost function or prediction horizon.
This can range from using a neural network to approximately predict the far look-ahead states in one step and predict an approximate end cost \cite{NeuralHorizon, Abdufattokhov} to learning the MPC value function \cite{Wabersich_valuefunctionLearning, Baltussen_ValueFuncApprox}.
In \cite{KRENERAdaptiveHorizon} the length of the prediction horizon is adjusted online and decreases as the system stabilizes.

In explicit MPC \cite{Bemporad_EMPC2009}, the implicit control law is replaced by a precomputed piecewise-affine approximation, avoiding online optimization entirely.
However, the number of regions and thus complexity  of the approach increases quickly with problem size \cite{Bemporad_EMPC2009}.

A similar idea is to design a robust MPC scheme such that a single control input can be used for a certain region of the state space.
For each region the control input can then be precomputed.
Thus, for online evaluation only the region and the corresponding control input has to be identified \cite{BayerTubebasedExplicit}.

Instead of splitting the state space into multiple regions, one can also use nonlinear function approximators such as neural networks or Gaussian processes to approximate the MPC law \cite{AKESSON2006937, LUKEN23_sobolev, TowingKiteslearnedmpc}. %
 In \cite{gonzalez2024neuralnetworksfastoptimisation}, the authors present an overview of different approaches to approximate a controller using neural networks. 
Obtaining closed-loop guarantees can be challenging with these approaches.
For linear systems, the authors of  \cite{Morari_ExplicitNN} show how to guarantee that the approximate controller produces feasible inputs.
In \cite{alsmeier2025imitationlearningmpcneural}, the authors show how to bound the approximation error of a neural network by choosing a sufficiently dense dataset.
The authors of \cite{Chatterjee_ExplicitquasiInterpol} propose a method to design an approximate controller, based on quasi interpolation, such that a given error margin is satisfied.
Other works mostly rely on statistical methods to verify the approximate controller for nonlinear systems \cite{KargProbValNNcontroller, HertneckAMPC}.

Few works exist with real world implementations of these approaches.
In \cite{Moennigmann_EmbeddedNN} a nonlinear autoregressive neural network is designed and employed on a microcontroller to control a small scale hydraulic plant.
The work in \cite{NubertNNAMPC_Robotmanipulator} approximates a robust MPC law using neural networks and applies it to control a robotic manipulator.
Slight parameter changes in the system model or MPC formulation usually require designing the approximate controller anew.
To this end, \cite{hose2024parameteradaptiveapproximatempctuning} exploit local sensitivity information to allow the approximate controller to adjust to parameter changes.
Their method is extended in \cite{hose2024finetuning_NNAMPC}, where Bayesian optimization is used to find suitable parameter of the MPC such that a higher objective function is optimized. They validated the approach experimentally.

While Gaussian process-based MPC has seen real-world applications for dynamics learning, to the best of our knowledge, no works apply Gaussian process approximations of the full MPC control law in embedded hardware scenarios for trajectory tracking.
GPs, however, offer several appealing properties for this purpose.
They can be refined incrementally online \cite{Kocijan_evolving_gps, Maiworm_onlineGPs}, which allows the controller to adapt when operating conditions change.
Moreover, although only the posterior mean is used in this work, the posterior variance provides a basis for safety-critical extensions.
Additionally, such uncertainty information could be incorporated into hierarchical control structures, as proposed in \cite{Koegel_hierarchical}, enabling an upper-level planner to account for prediction uncertainty.

We present a method to learn an efficient Gaussian process approximation of an MPC policy for trajectory tracking of an autonomous vehicle.
To this end, we reformulate the problem in curvilinear coordinates, which allows to efficiently represent the trajectory. 
Additionally, we use a feedforward control component to retain relevant structure.
To validate the applicability of the approximate controller, we employ it  on an embedded Raspberry Pi of a small test vehicle and compare its performance against an MPC implementation in ACADOS \cite{Verschueren2021ACADOS}, a tool specifically designed for fast MPC.

\section{Methods}

First, we outline model predictive control (MPC) for trajectory tracking.  We then introduce Gaussian processes (GPs) as a means to obtain an efficient approximation of the control law. Finally, we discuss challenges specific to approximating a trajectory-tracking controller.
\subsection{Model Predictive Control for Trajectory Tracking}
We consider a nonlinear discrete time system $ x_{k+1} = f(x_k,u_k,\eta_k)$,
with $ f: \mathbb{R}^{n_x} \times \mathbb{R}^{n_u} \times \mathbb{R}^{n_\eta} \rightarrow \mathbb{R}^{n_x}$
and where $x_k \in  \mathbb{R}^{n_x}$ are the states, $u_k \in  \mathbb{R}^{n_u} $ are the inputs and  $\eta_k \in  \mathbb{R}^{n_\eta} $ are the parameters of the system at time index $k \in \mathbb{N}$.
The goal is to steer the system along a reference trajectory $\{x^\text{ref}\}_0^{N_\text{cl}}=\{x_0^\text{ref},x_1^\text{ref},\hdots,x_{N_\text{cl}}^\text{ref}\}$
and $\{u^\text{ref}\}_0^{N_\text{cl}-1}=\{u_0^\text{ref},u_1^\text{ref},\hdots,u_{N_\text{cl}-1}^\text{ref}\}$ for all closed loop iterations $N_\text{cl} \in \mathbb{N}$.
To this end, we rely on MPC.
MPC repeatedly solves an optimal control problem (OCP) of the form
\begin{mini}
    {\{\hat{u}\}_0^{N-1}}{\sum_{i=0}^{N-1} \ell(\hat{x}_i,\hat{u}_i,x_{k+i}^\text{ref},u_{k+i}^\text{ref}) + E(\hat{x}_{N},x_{k+N}^\text{ref}),}{\label{eq:OCP1}}{}%
    \addConstraint{\hat{x}_{i+1} = f(\hat{x}_i,\hat{u}_i,\eta_{k+i})}{\quad \forall i \in \{0,1,\hdots,N-1\},}
    \addConstraint{g(\hat{x}_i,\hat{u}_i)\geq 0}{\quad \forall i \in \{0,1,\hdots,N-1\},}
    \addConstraint{\hat{x}_N\in\mathcal{X}_f}{,}
    \addConstraint{\hat{x}_0=x_k}{.}
\end{mini}

The goal is to find an optimal input sequence $\{\hat{u}^*\}_0^{N-1}$ and a corresponding sequence of states $\{\hat{x}^*\}_0^N$, which minimize the cost function with stage cost $\ell \! : \! \mathbb{R}^{n_x}\times \mathbb{R}^{n_u}\times \mathbb{R}^{n_x} \times \mathbb{R}^{n_u} \rightarrow \mathbb{R}_{\geq 0} $ and terminal cost $E \! : \! \mathbb{R}^{n_x} \times \mathbb{R}^{n_x} \rightarrow \mathbb{R}_{\geq0}$, while satisfying the constraints $g(x_i,u_i) \geq 0$ over the prediction horizon $N\in\mathbb{N}$.
In each iteration the first element of the optimal input sequence $u_k = \hat{u}^*_0$ is applied to the system.%

Thus, the control law $\mu$ is implicitly defined as the solution of the OCP, \ie $\mu(x,\Theta) = \hat{u}^*_0$, where we denote the tuple collecting all parameters, which can change from iteration to iteration, as $\Theta\in\mathbb{R}^{n_\Theta}$, where $n_\Theta \geq 0$ denotes the number of varying OCP parameters.
If we consider the OCP~\eqref{eq:OCP1}, in general, this tuple consists of the sequence of reference states, controls and model parameters, \ie $\Theta = \left(\{x^\text{ref}\}_k^{k+N},\, \{u^\text{ref}\}_k^{k+N-1}, \, \{\eta\}_k^{k+N-1} \right)$, leading to $n_\Theta = (N+1)n_{x}+N(n_{u} + n_\eta)$ parameters.
Consequently, due to the large amount of varying parameters, approximating the function $\mu(x,\Theta)$ efficiently is challenging.
Therefore, a core idea in this paper is to reformulate the problem in a way which reduces the number of parameters $n_\Theta$.

\subsection{Gaussian Process Regression}
In Gaussian process (GP) regression \cite{Rasmussen2006}, a non-parametric method, one considers a prior GP model $y_i = g(\xi_i)+\epsilon, \, \epsilon \sim \mathcal{N}(0,\sigma^2)$ and $g(\xi) \sim \mathcal{GP}(m(\xi),k(\xi,\xi'))$,
where $m(\cdot)$ is a mean function and $k(\cdot,\cdot)$ is a positive-definite kernel function.
These functions can be chosen by the designer of the GP. In this work, we rely on a zero mean prior $m(\xi)=0$ and, while other kernels can be used, a neural network kernel function
$    k(\xi,\xi') = s_f\arcsin{ \frac{\tilde{\xi}^T\Lambda \tilde{\xi}'}{\sqrt{(1+2\tilde{\xi}^\top \Lambda \tilde{\xi})(1+2\tilde{\xi}'^\top\Lambda \tilde{\xi}')} }}, $
where $\tilde{\xi} = [1 \, \xi^\top]^\top$.
We determine the hyperparameters $s_f\in\mathbb{R}$, $\Lambda \in \mathbb{R}$ by optimizing the logarithmic marginal likelihood \cite{Rasmussen2006}.
Given a data set $\mathcal{D} = \{ (\xi_i,y_i)\,|\,i=1,\ldots,n_D\}$ of features $\xi\in\mathbb{R}^{n_\xi}$ and labels $y\in\mathbb{R}$, we can compute the conditional distribution $g_*|Y \sim \mathcal{N}(m^+(\xi_*),k^+(\xi_*))$, with
\begin{subequations}
\begin{align}
    m^+(\xi_*) &= k_*^\top (K + \sigma^2 I)^{-1} Y \label{eq:postmean_full}, \\
    k^+(\xi_*) &=  k(\xi_*, \xi_*) - k_*^\top (K + \sigma^2 I)^{-1} k_*
\end{align}
\end{subequations}
to make predictions at unseen points $\xi_*$, where we introduced $K_{ij} = k(\xi_i,\xi_j), \, k_{*,i} = k(\xi_i,\xi_*)$ and $Y_i = y_i$ for ease of notation.
We base our approximation of the MPC law on \eqref{eq:postmean_full}.
To reduce computational complexity, we precompute $\alpha =  (K + \sigma^2 I)^{-1} Y $ in \eqref{eq:postmean_full} offline.
Thus online evaluation of the posterior mean \eqref{eq:postmean_full}  reduces to
\begin{align} \label{eq:postmean_fast}
    m^+(\xi_*) = k_*^\top\alpha.
\end{align}
Consequently, online computational complexity is linear with the amount of data $\mathcal{O}(n_D)$. 

\subsection{Approximate MPC for Trajectory Tracking}
The goal in approximate MPC is to design a computationally efficient mapping $\tilde{\mu}$, which closely represents the implicitly defined mapping $\mu$ of the MPC scheme.
In the trajectory tracking case the mapping $\mu$ not only depends on the state of the system but also on the set of parameters $\Theta$ at the current time step.%

Consequently, in order to approximate the mapping  sufficiently well, a straightforward approach is to use the initial condition $x$ as well as all parameters $\Theta$  as features $\xi\in\mathbb{R}^{n_\xi}$, with $n_\xi = n_x+n_\Theta$ of the approximating mapping.
As the parameters $\Theta$ contain the full-length reference trajectory this quickly leads to a high dimensional and thus typically hard and computationally expensive to approximate mapping. 
Therefore, we are motivated to keep the number of varying parameter $n_\Theta$ as low as possible.

In Section~\ref{subsec:ModelReform}, we  reduce this high dimensional mapping using a reformulation in curvilinear coordinates.
We further introduce a feedforward control part to the approximate mapping in Section~\ref{subsec:AMPC_car}, which allows to retain useful information.

\section{Approximate MPC for Autonomous Vehicle Trajectory Tracking}
In this section we introduce the vehicle model and its reformulation in curvilinear coordinates that simplifies the tracking problem.
Thereafter, we show how to learn a GP approximation of the MPC law.

\subsection{Kinematic Bicycle Model} \label{subsec:KBM}
We consider that the motion of the vehicle is given by the position $(p_x, p_y)$ and orientation $\psi$ of the vehicle in a planar plane, via a kinematic bicycle model, \ie  
$ \dot{p}_x = v\cos(\psi)$, $\dot{p}_y = v\sin(\psi)$, $\dot{\psi} = \frac{v\tan(\delta)}{L}$ and $\dot{v} = \frac{1}{T}(v_u-v)$.
Therein, $L$ is the distance between front and rear axle, $v$ is the velocity of the vehicle and $\delta$ denotes the steering angle. We express all states with the vector $\Tilde{x} = \left[p_x \, \, p_y \, \, \psi \right]^\top$ and controls with $\Tilde{u} = \left[v_u \, \, \delta\right]^\top$.
We expect the vehicle's low-level velocity controller to react to the velocity command with some time constant $T$.
After discretizing the model and formulating the OCP~\eqref{eq:OCP1}, the varying parameters in each step consist of the future reference states and controls.
For simplicity, we always assume a constant reference velocity, such that the parameters reduce to $\Theta = \left(\{\Tilde{x}^\text{ref}\}_0^{N}, \, \{\delta^\text{ref}\}_0^{N-1} \right)$ and thus $n_\Theta = 3(N+1)+1N$.

\subsection{Model Reformulation}\label{subsec:ModelReform}
To limit the number of varying parameters $n_\Theta$ and thereby reduce the GP’s feature dimension, we adopt an alternative model formulation.
In this formulation, the full-length reference trajectory is condensed into a constant reference state, supplemented by a single varying parameter that captures its essential information.
We do so by transforming the model into curvilinear coordinates around the reference path, similar to  \cite{KLOESER2020_PathparametricModel, STEINKE22_motionsickness}
\begin{subequations}\label{eq:VehicleModel}
\begin{align}
    \dot{s} &= \frac{v\cos(\alpha)}{1-n\kappa^c(s)}, \label{eq:dot_s}\\
    \dot{n} &= v\sin(\alpha), \label{eq:dot_n}\\
    \dot{\alpha} &= \frac{v}{L}\tan{\left( \delta \right)} - \kappa^c(s)\dot{s}, \label{eq:dot_alpha}\\
    \dot{v} &= \frac{1}{T}(v_u-v) \label{eq:velocity_control}.
\end{align}
\end{subequations}

In \eqref{eq:VehicleModel}, the control input remains unchanged $u = \Tilde{u} = \left[v_u \, \,  \delta\right]^\top$. %
 The new state vector is $x = \left[s \, \, n \, \, \alpha \, \, v\right]^\top$.
Here, we denote the distance of the vehicle to the reference path with $n$.
The difference between the orientation of the vehicle to the corresponding tangent line on the reference is $\alpha$.
The curvature of the reference path $\kappa^c$ is parameterized by the arc length parameter $s$.
Ideally, we can describe the reference curvature $\kappa^c(s)$  exactly.
However, in more realistic situations, where the track is not easily describable by such a function, we use an approximation of the true track curvature.
In \cite{KLOESER2020_PathparametricModel}, this function is for example modeled using third-order B-splines.
For practical realization, we propose to  rely on a look-up table and linearly interpolate between two consecutive curvature values.

In this reformulation, following the desired track is mostly described by a constant state, \ie $x_\text{ref} = \left[s_\text{ref}\,\, n_\text{ref}\,\, \alpha_\text{ref} \,\, v_\text{ref} \right]^\top = \left[\star\, \, 0 \, \, 0 \, \, r_v\right]^\top$, where the symbol  $\star$ indicates that we allow the arc length parameter to increase indefinitely as the vehicle progresses along the track.

Regarding the velocity, we assume that the goal is to drive with a constant velocity along the track, \ie  $v_\text{ref} = r_v >0 $ and $v_{u,\text{ref}}=r_v$.
The steering angle needed to stay on the track is $\delta_\text{ref}(\kappa^c(s)) = \arctan\left(L\kappa^c(s)\right)$.
Overall, we therefore have 
$u_\text{ref} = \left[r_v \, \,  \delta_\text{ref}(\kappa^c(s))\right]^\top$.

Note that the curvature 
$\kappa^c(s)$ encodes both the geometric shape of the reference path and the corresponding steering angle $\delta_\text{ref}$.
To allow a generalization to varying tracks we can therefore directly use the track curvature information, when learning the controller.
Overall, rather than describing the reference by its full two-dimensional position, orientation, and steering profiles, we can now describe the reference in terms of a single parameter $\kappa^c$.
After discretization of \eqref{eq:VehicleModel}, we obtain a model in form of
\begin{align} \label{eq:discreteSystem_general}
     x_{k+1} = f(x_k,u_k, \kappa^c_k),
 \end{align}
where $k$ is the discrete time index such that the current continuous time is $t = kT_s$ and $T_s$ is the sampling time.
Therefore, the parameters of the system are now $\Theta = \{\kappa^c\}_k^{k+N-1}$.
Consequently, we reduced $n_\Theta = 3(N+1)+N$ in Section~\ref{subsec:KBM} to $n_\Theta = N$.

\subsection{Design of the GP Approximation} \label{subsec:AMPC_car}

To retain structure present in the control law and avoid discarding information that can be expressed analytically, we introduce a feedforward component that captures the nominal input corresponding to curvature tracking.
As described in Section~\ref{subsec:ModelReform}, the input that keeps the system on the reference path is given by
\begin{align}\label{eq:u_ff}
    u_\text{ff}(\kappa^c) = \begin{bmatrix}
        r_v \\ \arctan(L\kappa^c)
    \end{bmatrix}.
\end{align}
We further define the residual control input as $\Delta u = \hat{u}^*_0(x,\kappa^c) - u_\text{ff}(\kappa^c)$.
Instead of learning the mapping to the optimal control input directly, we therefore propose to learn the residual control input, \ie $y = \Delta u$.
Consequently, during online operation, the control input applied to the system is reconstructed as
\begin{align}\label{eq:approximatecontrolaw}
    \hat{u} =u_\text{ff}(\kappa^c)+\bar{m}^+(\xi),
\end{align}
where $\bar{m}^+$ is the vector of the posterior means~\eqref{eq:postmean_fast} of the designed GPs.
That is, we design a separate GP for each dimension of the control input.

Since the computation complexity of the GP prediction \eqref{eq:approximatecontrolaw} scales linearly with the amount of data, identifying promising points helps reducing online computational complexity.
To this end, we propose to adopt an approach similar to \cite{Mesbah_offsetfree_learning} and \cite{Rose_reachable}.
That is, we first define a set of possible initial conditions $\xi\in \Xi$ in the feature space. 
We then approximately compute the closed loop reachable set using the sampling based approach outlined in \cite{Rose_reachable}.
We restrict our search for the most relevant data points to this space.
Finally, we greedily include new data points $\xi^*$ in the GP, which maximize the objective function, \ie
\begin{align}\label{eq:greedy_data_selection}
    \xi^* = \arg\max_{\xi\in\Xi} \norm{y(\xi)- \bar{m}^+(\xi)}.
\end{align}
Solving \eqref{eq:greedy_data_selection} is challenging as $y(\xi)$ is implicitly defined through solving the OCP~\eqref{eq:OCP1}. 
We therefore solve \eqref{eq:greedy_data_selection} approximately, by sampling points in $\Xi$ and select the sample which maximizes \eqref{eq:greedy_data_selection}.

\section{Experimental Validation}
In this section we present the experimental evaluation of the proposed GP-based approximate controller.
To this end, we first present the  OCP. 
Thereafter, we design the GPs to approximate the MPC law.
We then assess the computational efficiency of the controller in comparison to the MPC implementation generated with \mbox{ACADOS}.
Finally, we demonstrate its closed-loop performance on a small-scale autonomous vehicle.

\subsection{Resulting Optimization Problem}\label{subsec:OCP_car}
For the OCP~\eqref{eq:OCP1} we use \eqref{eq:discreteSystem_general} to model the vehicle dynamics.
For the stage cost we use $\ell(x,u,x^\text{ref},u^\text{ref}) = \ell_x(x,x^\text{ref}) + \ell_u(u,u^\text{ref})$ with $\ell_x(x,x^\text{ref}) = w_1(nv)^2+w_2 \alpha^2 + w_3 (v-r_v)^2$, $\ell_u(u,u^\text{ref}) = (u-u_\text{ref})^\top W_u (u-u_\text{ref})$
 and corresponding weights $w_i > 0, \, \forall i = {1,2,3}$ as well as $W_u >0 $. Table~\ref{tab:MPCparas} shows all parameter values used in the optimal control problem~\eqref{eq:OCP1}.
The reference velocity is $r_v = \SI{0.5}{\meter\per\second}$.
 We do not penalize the arc length parameter $s$, because it is allowed to grow until the reference ends.
At the end of the reference, we reset $s=0$,  since the vehicle follows a circular track.
Also note that we penalize the product of the velocity times the distance to the reference in $\ell_x$. This way we penalize the covered area which the vehicle is off the reference and thus larger deviations are more strongly weighted at higher speeds. %
For the end cost in \eqref{eq:OCP1}, we use $E(x_N,x^\text{ref}_N) = \ell_x(x_N,x^\text{ref}_N)$.

We do not impose constraints on the states, but on the inputs.
The velocity is constrained to a positive value less than $\SI{1.2}{\meter\per\second}$, \ie $0\leq v \leq 1.2$.
We constrain the steering angle to be between $-12\frac{\pi}{180} \,\si{\radian} \leq \delta \leq 12\frac{\pi}{180}\,\si{\radian}$,  as above and below these angles the driven curvature of the vehicle remained the same during experiments.

\begin{table}[]
\caption{Parameters for the Optimal Control Problem \eqref{eq:OCP1}.}\label{tab:MPCparas}
\centering
\begin{tabular}{l|l}
 Parameter & Value                       \\
  \hline
$\left[ w_1, \, w_2, \, w_3\right]$ & $\left[ 100, \, 5, \, 5 \right]$   \\
$W_u$ & $\text{diag}(\left[5, \, 2\right])$   \\
$N$ & 40 \\
$\left[ v_\text{min}, \, v_\text{max}\right]$ & $\left[ 0, \, 1.2 \right]$ \\
$\left[ \delta_\text{min}, \, \delta_\text{max}\right]$ & $\left[ -12\frac{\pi}{180} \,\si{\radian}, \, 12\frac{\pi}{180} \,\si{\radian} \right]$\\
$L$ & $\SI{0.16}{\meter}$\\
$T_s$ & \SI{0.01}{\second}\\
$T$ & \SI{0.1}{\second}
\end{tabular}
\end{table}

\subsection{Approximate GP Controller for the Vehicle}
As outlined in the preceding Section~\ref{subsec:ModelReform}, the reference trajectory information is now fully represented by the curvature profile over the prediction horizon, \ie $\Theta = \{\kappa^c\}_k^{k+N-1}$.
To simplify the representation further, we propose to restrict the curvature information to only the first  $N_\kappa \leq N-1$ elements, \ie $\Theta = \{\kappa^c\}_k^{k+N_\kappa}$.
This truncation is justified by the observation that future curvature values have only a negligible influence on the first optimal control action.
Throughout this work, we set $N_\kappa = 0$. That is we only provide information about the local reference curvature. With this choice, the number of varying parameters further decreases  to $n_\Theta = 1$. 
Accordingly, the feature vector of the approximate controller becomes $\xi = \left[n \, \, \alpha \, \, v\, \, \kappa^c\right]^\top$.

\begin{figure}
    \centering
\includegraphics[]{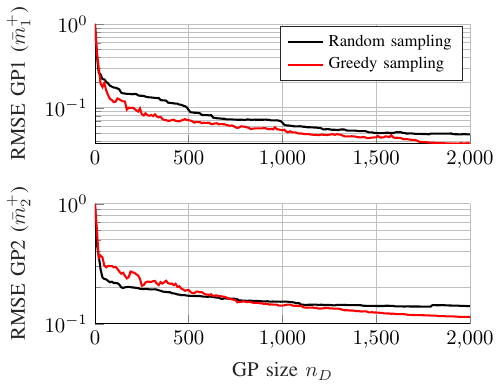}%
    \caption{Root mean squared error (RMSE) for $3000$ unseen test data of each GP depending on the number of data points $n_D$.}%
    \label{fig:GP_RMSE_size}
\end{figure}

By repeatedly solving~\eqref{eq:greedy_data_selection}, we iteratively add $2000$ data points greedily to each GP.
We depict the root mean squared error for $3000$ unseen points, depending on the number of active data in Figure~\ref{fig:GP_RMSE_size}.
 Additionally, we show a comparison to a design, with randomly selected data points.
For the first GP, greedily selecting points strictly outperforms a random selection of points. For the second GP, greedy selection outperforms random selection in the long run. 
For the embedded controller we  use $n_D = 1000$ active data points, which provides a good trade-off between accuracy and computational effort.%

\subsection{Computational Efficiency}
To assess the computational efficiency of the proposed GP-based approximate controller, we benchmark the average execution time required to compute a control input and compare it against an MPC implementation generated with ACADOS \cite{Verschueren2021ACADOS} using the real-time iteration (RTI) scheme. Both controllers were executed on a desktop machine with an Intel Core i7-1165G7 CPU (4 cores, 32 GB RAM) using MATLAB’s timeit() function.
The results are depicted in Figure 2. The ACADOS-based controller exhibits a median computation time of approximately $\SI{2.7e-4}{\second}$, while the GP-based controller achieves a median of $\SI{5e-5}{\second}$, corresponding to a speedup factor of approximately five.

\begin{figure}
    \centering
  \includegraphics[]{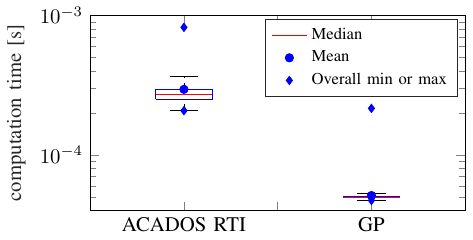}%
    \caption{Computation times of ACADOS and the GP. The blue box covers the $25$ to $75$ quantile, while the whiskers extend to the non outlier minimum and maximum. Values outside of  $1.5$ times the interquantile range are considered as outliers. For clarity of the figure, instead of all outliers, only the overall minimum and maximum computation times are depicted with blue diamonds. The median is indicated by the red line, while the blue dot indicates the mean computation time.}%
    \label{fig:comptimes_statistics}
\end{figure}

\subsection{Performance of the Embedded Approximate Controller}
We use the code generation functionality of MATLAB to generate C code of the approximating controller \eqref{eq:approximatecontrolaw}, which is then employed on the embedded Raspberry Pi of the vehicle for online control. %
The embedded controller is run with a sampling time of $\SI{1}{\milli\second}$ on the vehicle.
As comparison, we run ACADOS on an external PC with a sampling time of $\SI{5}{\milli\second}$ to compute the optimal control input and drive the vehicle around the same track.

\begin{figure}
    \centering
    \includegraphics[]{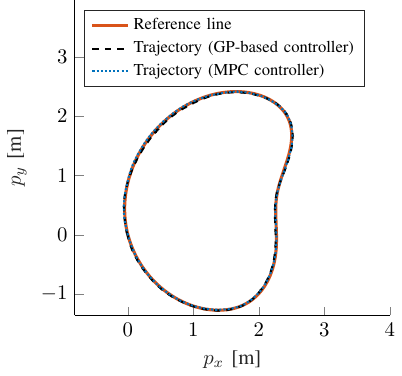}%
    \vspace{-3mm}
    \caption{Reference track and driven paths by the MPC and GP controlled vehicle.}%
    \label{fig:ref_track_andcurv}
\end{figure}

\begin{figure}
    \centering
   \includegraphics[]{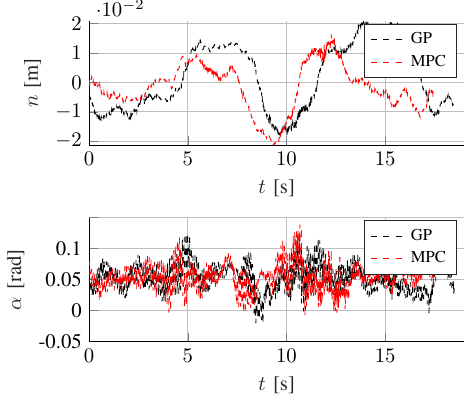}%
    \vspace{-3mm}
    \caption{Closed loop states during a single circuit. The black dashed line represents the GP controlled system, while the system controlled by the MPC is depicted with the red dashed line.}
    \vspace{-3mm}
    \label{fig:cl_states}
\end{figure}

In Figure~\ref{fig:ref_track_andcurv}, we show the reference track as well as the driven paths when using both controller.
In Figure~\ref{fig:cl_states}, we depict the measured states $n$, and  $\alpha$ during one lap by employing the MPC controller as well as its approximation.
The deviation from the reference path remains below $\SI{2}{\centi\meter}$ and the heading deviation $\alpha$ are similar for both controller. 
To quantify the performance of both controller, we compute their closed loop cost according to $J_\text{cl} = \frac{1}{N_\text{cl}} \sum_{k=0}^{N_\text{cl}-1} \ell(x_k,u_k,x^\text{ref}_k,u^\text{ref}_k)$, where $N_\text{cl}$ indicates the length of the closed loop trajectory.
For the embedded controller we obtain a cost of $0.0092$, while the MPC has a cost of $0.0081$. Therefore, both controller have a similar closed loop cost.

\pagebreak\section{Conclusion}

Although model predictive control is a powerful approach for controlling complex systems, its high computational demands often make real-time implementation on embedded devices infeasible. Using a Gaussian process to approximate the model predictive control law eliminates the need for solving nonlinear optimization problems online, significantly reducing computational overhead. We develop a Gaussian process-based approximation for trajectory tracking of an autonomous vehicle. By reformulating the problem in curvilinear coordinates, the reference trajectory is represented efficiently. 

The resulting controller is deployed on a Raspberry Pi and experimentally validated on a small-scale vehicle. Compared to an implementation in ACADOS, the Gaussian process-based controller achieves a computational speedup of over five times with comparable closed-loop performance. 

Future work will focus on exploiting uncertainty estimates that Gaussian processes provide to guide online refinement or enhance safety.

\section*{Acknowledgment}%
The authors thank former student Yang Yin for the discussions, which contributed to the ideas in this paper.

\bibliographystyle{IEEEtran}
\bibliography{literature}

\end{document}